\documentclass[pra,nofootinbib,reprint]{revtex4-1}

\usepackage{amsmath}
\usepackage{color}
\usepackage{comment}
\usepackage{graphicx}
\usepackage{xspace}

\graphicspath{{../fig/}{./}}

\newcommand{\dd}{\ensuremath{\mbox{d}}\xspace}
\newcommand{\jnt}[4]{\ensuremath{\int\limits_{#1}^{#2}\, {#3}\,\dd {#4}}}
\newcommand{\intinf}[2]{\jnt{-\infty}{+\infty}{#1}{#2}}
\newcommand{\res}{\ensuremath{_{\mbox{\tiny R}}}}
\newcommand{\fin}{\ensuremath{_{\mbox{\tiny end}}}}
\newcommand{\rfr}{\ensuremath{^{\mbox{\tiny (ref)}}}}
\newcommand{\sca}{\ensuremath{_{\mbox{\tiny sca}}}}
\newcommand{\cpl}{\ensuremath{_{\mbox{\tiny cpl}}}}
\newcommand{\oned}{\ensuremath{\mbox{1\sc d}}\xspace}

\newcommand{\acronym}[1]{{\mbox{\sc #1}}\xspace}
\newcommand{\lift}{\acronym{lift}}

\newcommand{\rabbit}{\acronym{rabbit}}

\newcommand{\tdse}{\acronym{tdse}}

\newcommand{\dos}{\acronym{dos}}

\newcommand{\pes}{\acronym{pes}}
\newcommand{\PES}{\pes}
\newcommand{\tpes}{\ensuremath{\mbox{\acronym{tpes}}}\xspace}
\newcommand{\tPES}{\tpes}

\newcommand{\wX}{\ensuremath{\omega_{\mbox{\small xuv}}}\xspace}
\newcommand{\tauX}{\ensuremath{\tau_{\mbox{\small xuv}}}\xspace}
\newcommand{\ket}[1]{\ensuremath{\left\vert#1\right\rangle}}

\definecolor{jcol}{rgb}{1,0,0}
\definecolor{jcol}{rgb}{0,0,0}
\newcommand{\jeremie}[1]{\textcolor{jcol}{#1}\xspace}

\newcommand{\W}{\ensuremath{\mathcal{A}}\xspace}
\newcommand{\PEt}{\ensuremath{\mathcal{P}(E,t)}\xspace}
\newcommand{\G}{\ensuremath{\mathcal{G}}\xspace}
\newcommand{\E}{\ensuremath{\mathcal{E}}\xspace}
\newcommand{\F}{\ensuremath{\mathcal{F}}\xspace}

\newcommand{\VE}{\ensuremath{V_{\bar E}}\xspace}
\newcommand{\A}{\ensuremath{A(E)}\xspace}

\begin{document}
\title{Ionisation dynamics through a Fano resonance: \\ Time domain interpretation of spectral amplitudes}
\author{Antoine Desrier}
\author{Alfred Maquet}
\author{Richard Ta\"\i eb}
\author{J\'er\'emie Caillat}
\email{jeremie.caillat@sorbonne-universite.fr}
\affiliation{Sorbonne Universit\'e, CNRS, Laboratoire de Chimie Physique-Mati\`ere et Rayonnement, LCPMR, F-75005 Paris, France}
\date{\today}
\begin{abstract}
We investigate a conjecture used in recent experiments to reconstruct the complete dynamics of Fano autoionisation processes out of measured spectral amplitudes [Gruson {\em et al}, Science {\bf 354} 734 (2016); Beaulieu {\em el al}, Science {\bf 358} 1288 (2017); Busto {\em et al}, J. Phys. B: At. Mol. Opt. Phys. {\bf 51} 044002 (2018)]. The validity of the conjecture is established analytically within the formalism of Fano, and tested numerically on model atoms displaying adjustable autoionising states. A general condition for which the conjecture is valid, beyond the Fano case, is then derived.
\end{abstract}
\maketitle



\subsection{Introduction}

Amongst the variety of phenomena explored in the time domain by attosecond (1  as = 10$^{-18}$ s) time-resolved spectroscopy~\cite{leone2014a,calegari2016a}, a very fundamental one is photoemission itself. 
 In the late 2000's, cutting edge experiments performed in atoms~\cite{schultze2010a,klunder2011a}, molecules~\cite{haessler2009a} and solids~\cite{cavalieri2007a} evidenced for the first time ultra short ``ionisation delays''. These are equivalent to  Wigner-Smith group delays~\cite{wigner1955a} applied to photoemission: They reflect how  the  dynamics of a photoelectron  is affected by species- and channel-specific short range interactions with the composite parent ion, and are encoded in the scattering phase of the associated wave-functions. 
Photoemission group delays are  accessed experimentally through interferometric measurements of photoemission amplitudes, and more specifically of their phases~\cite{yakovlev2010a,dahlstrom2012a,pazourek2015a,cattaneo2016a}. 

The relevance of mere ``delays'' to accurately characterise the  dynamics of photoemission is however restricted to smooth continuua displaying no or little structure, where the scattering phase varies linearly within the bandwidth of the photoelectron wave packets. 
Reducing photoemission dynamics to a single group delay no longer applies when the wave packet builds up across a significantly structured continuum. This is typically the case for a Fano resonance -- a photoemission process taking place competitively through an autoionising metastable state and directly to the continuum~\cite{fano1961a}. 

It is only recently that comprehensive ways of looking at autoionisation in the time domain have been proposed~\cite{zhao2005a,wickenhauser2005a,mercouris2007a,mercouris2013a,chu2010a,granadoscastro2013a,yang2018a,busto2018a}, motivated by the perspectives of probing and controlling these dynamics experimentally with unprecedented, sub-femtosecond, resolution~\cite{gilbertson2010a,ott2013a,lin2013a,kotur2016a}. Complete Fano resonance build-ups were monitored experimentally for the first times using an original interferometric scheme dubbed Rainbow-\rabbit~\cite{gruson2016a,busto2018a} and with attosecond transient absorption spectroscopy~\cite{kaldun2016a}, both on the He($2s2p$) prototype~\cite{fano1961a}. An extension of the Rainbow-\rabbit technique was  used shortly after to reconstruct the anisotropic, polarisation sensitive, photoemission dynamics of chiral molecules in a smooth continuum as well as in the vicinity of an autoionising state~\cite{beaulieu2017a}. 

The present work addresses the reconstruction method introduced in~\cite{gruson2016a}, based on a time-energy analysis of the scattering amplitudes $A(E)$ measured around a Fano resonance\footnote{The possibility to {\em measure} the phase of a photoemission amplitude  using pump-probe interferometry has been thoroughly investigated over the past years~\cite{dahlstrom2012a,yakovlev2010a,zhang2011a,jimenez2014a,jimenez2016a,cattaneo2016a}. It is still the subject of open questions (see {\em eg}~\cite{heuser2016a,lindroth2017a,vacher2017a}) that we do not address in the present paper, which instead focuses on the time-domain {\em interpretation}.}. In spite of being appreciably straightforward, the experimental data treatment relies on a {\em conjecture} consisting in giving sense to  a ``time-domain amplitude'' defined as
\begin{eqnarray}\label{eqn:E2t}
a(t)&:=&\intinf{\A e^{-iE t}}{E}.
\end{eqnarray}
Switching between the spectral and time domains using Fourier relations is almost a reflex inherited from wave mechanics. In quantum physics, it applies to {\em wave-functions} as prescribed by the time-dependent Schr\"odinger equation. However, when applied to {\em amplitudes} as in Eq.~\ref{eqn:E2t} the validity of time domain interpretations is not granted by the first principles of quantum mechanics. In this context, the aim of this work is to establish the conditions under which $a(t)$ is a physically meaningful quantity, first in the specific case of Fano autoionisation and then in a broader perspective.

The paper is organised as follows. First, we specify in Section~\ref{sec:Fanoconjecturepres} the above mentioned conjecture within the formalism established by Fano. Then, in Section~\ref{sec:Fanodyn}, we demonstrate analytically that the conjecture is indeed valid for any Fano autoionisation process. In Section~\ref{sec:numerics}, we  illustrate the analytical result and test the main approximation used in their derivation by means of numerical simulations performed on simple model atoms. Finally, we derive a very general condition under which the conjecture is valid, beyond the Fano case, in Section~\ref{sec:CGV}. The conclusions are given in Section~\ref{sec:conc}. Technical details on the analytical derivations and numerical simulations are provided in appendices.

Atomic units are used in Eq.~\ref{eqn:E2t} and  all through the article unless otherwise stated.

\subsection{Fano spectral amplitudes and their conjectured time-domain interpretation}\label{sec:Fanoconjecturepres}
In this section, we first provide a short account of the formalism developped by Fano to model autoionising resonances in the spectral domain, and then expose the conjectured time-domain interpretation of spectral amplitudes discussed in this paper.

\subsubsection{Fano's formalism: autoionisation in the spectral domain}\label{sec:Fanoparadigm}

The formalism established by Fano~\cite{fano1961a} to describe autoionisation in the spectral domain uses a partitioning of the system's eigenstates  (\ket{\Psi_E}) at energies $E$ around the resonance in terms of {\em bound} ($\ket{\phi_{\mbox{\tiny b}}}$)  and {\em scattering} ($\ket{\jeremie{\phi}_E}$) contributions\footnote{To keep the notations simple, we discard the additional parameters that, besides the energy $E$, allow to fully specify the considered  continuum states and therefore to distinguish the various possible ionisation channels.},
\begin{eqnarray}\label{eqn:psires}
\ket{\Psi_E}&=&\ket{\phi_{\mbox{\tiny b}}}+\ket{\jeremie{\phi}_E},
\end{eqnarray}
each of them being assigned appropriate asymptotic behaviors.

The bound part typically corresponds to a doubly excited state (such as the 2$s$2$p$ configuration of He), while the scattering part is a relaxed singly ionised state with an electron in the continuum [He(1$s$)+e$^-$].   
Any population of $\ket{\phi_{\mbox{\tiny b}}}$ eventually transfers into $\ket{\jeremie{\phi}_E}$ due to electron correlation, within a time scale given by the resonance lifetime.  It is the coherence of the process which results in interferences between the two paths, and shape the early times of the correlation driven scattering dynamics of the photoelectron. 

\jeremie{The approach further consists in expanding the continuum components \ket{\phi_E} in terms of reference continuum states \ket{\varphi_E}  virtually uncoupled from the bound component \ket{\phi_{\mbox{\tiny b}}}. 
} For a resonance with characteristic energy $E\res$ and width $\Gamma\res$, Fano derived the following expression for the probability amplitude \A to end-up in state \ket{\varphi_E} upon absorption of a photon with frequency  $\omega=E+I_p$ ($I_p$ is the ionisation potential):
\begin{eqnarray}\label{eqn:AEFano}
\A &=&\mathcal{F}(\omega) V_E \frac{\varepsilon+q}{\varepsilon+i}.
\end{eqnarray}
In this compact formula,   $V_E$ is the direct transition amplitude from the initial bound state \ket{\phi_0} towards \ket{\varphi_E}, $\varepsilon=2(E-E\res)/\Gamma\res$ is the reduced energy and the Fano parameter $q$ is proportional to the ratio $V_E/V_{\tiny\mbox{b}}$, where  $V_{\tiny\mbox{b}}$ is the transition amplitude from \ket{\phi_0} towards \ket{\phi_{\mbox{\tiny b}}}, see~\cite{fano1961a} for more details. In order to take into account finite pulse effects, we have included in this expression the driving field amplitude $\mathcal{F}(\omega)$. 
As evidenced by Eq.~\ref{eqn:AEFano}, the scattering phase  $\arg \A$ undergoes two $\pi$ jumps within a spectral range of  $\sim\Gamma\res$ around $E\res$: A sharp one occurs at $\varepsilon=-q$ as the numerator of \A vanishes and changes sign, and another one, smoothed over the resonance width, is centered on the resonance energy as the real part of the denominator vanishes and changes sign. 

The essential approximation made by Fano to derive Eq.~\ref{eqn:AEFano} is to consider the direct transition amplitude towards the scattering continuum, $V_E$, to be constant within a range of a few $\Gamma\res$ in the vicinity of the resonance. This is a very reasonable approximation, as highlighted by the remarkably broad efficiency of the Fano profile to model with high accuracy resonances in nuclear, atomic, molecular and nano physics, and even classical oscillators, see for exemple the review~\cite{miroshnichenko2010a}.  As we will see in section~\ref{sec:Fanodyn}, it is also the only approximation needed to validate the time domain interpretation of $A(E)$ made in~\cite{gruson2016a,beaulieu2017a,busto2018a}.

\subsubsection{The conjecture: autoionisation in the time domain}
In the time domain, the most straightforward characteristic of a Fano resonance is its lifetime, which coincides with the ionisation delay {\em at resonance}. It is however only representative of the exponential decay of the autoionising state into the continuum, discarding the impact of interferences on early times dynamics.  More detailed temporal insight on resonances can be obtained by considering the ``survival probability'' introduced by Krylov and Fock~\cite{krylov1947a} 
. 
It notably displays deviations from the exponential law at small and large times
~\cite{fonda1978a}. Still, it focuses on the resonance {\em decay}, considering that only the bound part of the resonance is initially populated. It may be suitable for describing an Auger or a nuclear decay but excludes other processes such as Fano resonances. 

In contrast, the approaches based on the Rainbow-\rabbit technique~\cite{gruson2016a,beaulieu2017a,busto2018a}  mentioned in the introduction provide a whole picture of the Fano resonance as builds up in time.
The conjecture it relies on consists in relating the temporal amplitude $a(t)$, defined  in Eq.~\ref{eqn:E2t}, to two observables: the ionisation rate, $I(t)$, according to 
\begin{eqnarray}\label{eqn:It}
I(t)&=&\frac{1}{\jeremie{2\pi}}\vert a(t)\vert^2,
\end{eqnarray}
and the photoelectron spectrum as it builds up in time during the process, $\PEt$, according to
\begin{eqnarray} \label{eqn:PEt}
\PEt&=& \frac{1}{4\pi^2} \left\vert \jnt{-\infty}{t}{a(t')e^{iEt'}}{t'} \right\vert^2.
\end{eqnarray}
The latter is hereafter referred to as the {\em transient photoelectron spectrum} (\tPES). 

The two quantities being related through
\begin{eqnarray}
I(t)&=&\frac{\partial}{\partial t}\intinf{\PEt}{E},
\end{eqnarray}
it is easy to show that Eq.~\ref{eqn:PEt}, when valid, implies Eq.~\ref{eqn:It} (see Appendix~\ref{sec:demo1}).  We will thus focus on Eq.~\ref{eqn:PEt}. This puts forward the limited inverse Fourier transform (\lift) of $a(t)$, defined as  
\begin{eqnarray}\label{eqn:WEtdef}
\W(E,t)&:=&\frac{1}{2\pi}\jnt{-\infty}{t}{a(t')e^{iEt'}}{t'},
\end{eqnarray} 
which is the main tool inherited from time-frequency analysis in the present study. The $(2\pi)^{-1}$ normalisation factor \jeremie{in Eq.~\ref{eqn:WEtdef}} 
is here to ensure consistency between the Fourier transform and its inverse, throughout the article.

\subsection{Analytical validation of the conjecture}\label{sec:Fanodyn}

In this section, we  demonstrate analytically that the conjectured interpretation of $a(t)$ is indeed valid for a Fano resonance. To this end, we compare  the dynamics {\em inferred} from  $a(t)$ to the {\em actual} Fano dynamics as initially derived by Mercouris and co-workers~\cite{mercouris2007a,mercouris2013a}.  

\subsubsection{Actual build-up of the continuum wave packet}

In  the framework provided by Fano's approach, the evolution of the {\em continuum} wave packet \ket{\psi\sca(t)} during its formation and propagation is obtained by \jeremie{expanding it on the reference scattering states \ket{\varphi_E}},  
\begin{eqnarray}\label{eqn:ewp}
\ket{\psi\sca(t)}&=&\intinf{c_E(t)\ket{\varphi_E}e^{-iEt}}{E}.
\end{eqnarray}
The \tPES is  directly related to the expansion coefficient $c_E(t)=\langle E \vert \psi\sca(t) \rangle$ through
\begin{eqnarray}\label{eqn:PEtdef}
\PEt&=&\vert c_E(t) \vert^2.
\end{eqnarray}
Therefore the conjecture~\ref{eqn:PEt} is verified if and only if the modulii of $\W(E,t)$ (defined in Eq.~\ref{eqn:WEtdef}) and of $c_E(t)$ are equal.

A comprehensive derivation of the analytical expression for $c_E(t)$ is detailed in~\cite{mercouris2007a,mercouris2013a}\footnote{A similar expression is derived in~\cite{chu2010a} for a Fano resonance triggered by a {\em sudden pulse}.}.
It reads
\begin{eqnarray} \nonumber
c_E(t)&=&\frac{V_E}{i} 
\left[\G(E,t) \frac{(q+\varepsilon)}{(\varepsilon+i)}
\right. \\ && \left. 
-\G(E\res-i\frac{\Gamma\res}{2},t)\frac{(q-i)}{(\varepsilon+i)}e^{i\varepsilon\frac{\Gamma\res}{2}t}e^{-\frac{\Gamma\res}{2}t}\right]
\label{eqn:cEtmerc}
\end{eqnarray}
where
\begin{eqnarray}\label{eqn:GEt}
\G(E,t)&=&\frac{1}{2\pi}\jnt{-\infty}{t}{\mathcal{E}(t')e^{i(E+I_p)t'}}{t'}
\end{eqnarray}
\ \\ \\ is the \lift of the field temporal amplitude
\begin{eqnarray}
\E(t)&=&\intinf{\F(\omega)e^{-i\omega t}}{\omega}.
\end{eqnarray}
One can easily verify with Eqs.~\ref{eqn:cEtmerc} and~\ref{eqn:AEFano}  that the asymptotic value of $c_E(t)$ is proportional to the transition amplitude $\A $,
\begin{eqnarray}\label{eqn:AEcEt}
\lim\limits_{t\rightarrow+\infty}c_E(t)&=&\frac{1}{i}\A ,
\end{eqnarray}
a result  reminiscent of the perturbative treatment underlying the Fano formalism.

\subsubsection{Reconstruction of the continuum wave packet}

We now derive the analytical expression of $\W(E,t)$ for a Fano resonance, that we will compare to  the one of the time-dependent coefficients $c_E(t)$ given in Eq.~\ref{eqn:cEtmerc}. Assuming that $V_E$ is constant ($=\VE$) over the bandwidth of \A, one can show that the  temporal amplitude (Eq.~\ref{eqn:AEFano}) for a Fano resonance verifies\footnote{The dynamics of atomic autoionisation with a sudden pulse was also investigated theoretically in~\cite{zhao2005a}. In this paper, the conjecture corresponding to Eq.~\ref{eqn:It} is implicit  (see Eq. 2 of that reference). Their results are compatible with the present expression  for $a(t)$ (Eq.~\ref{eqn:atFano}) in the limit  $\mathcal{E}(t)\rightarrow\delta(t)$.}
\begin{eqnarray}\label{eqn:atFano}
\frac{1}{\VE}
 a(t)
&=&\mathcal{E}(t)e^{I_pt}+(q-i) \frac{\Gamma\res}{2}
\nonumber \\ && \times
\intinf{\frac{\mathcal{F}(\omega')}{\omega'-\omega'\res+i\frac{\Gamma\res}{2}}e^{-i(\omega'-I_p) t}}{\omega'}. 
\end{eqnarray}
Therefore the \lift of $a(t)$, defined in Eq.~\ref{eqn:WEtdef}, reads
\begin{widetext}
\begin{eqnarray}	
\frac{1}{\VE}
\W(E,t)&=& \nonumber
\G(E,t)+(q-i)\frac{\Gamma\res}{2}
\jnt{\omega'=-\infty}{+\infty}{
	\frac{\mathcal{F}(\omega')}{\omega'-\omega\res+i\frac{\Gamma\res}{2}}
	\frac{1}{2\pi}
	\jnt{t'=-\infty}{t}{
		e^{i(\omega-\omega') t'}
	}{t'}
}{\omega'} \\
&=&
\G(E,t)-(q-i)\frac{\Gamma\res}{2}\frac{i}{2\pi}\intinf{
\frac{\mathcal{F}(\omega')}{(\omega'-\omega\res+i\frac{\Gamma\res}{2})}\lim\limits_{t_0\rightarrow-\infty}\frac{\left[e^{i(\omega-\omega') t}-e^{i(\omega-\omega') t_0}\right]}{(\omega'-\omega)}
}{\omega'}. \nonumber \\ 
\end{eqnarray}
\end{widetext}
In order to get a formula involving \G functions in each term, as in Eq.~\ref{eqn:cEtmerc}, we reintroduce the pulse temporal profile $\mathcal{E}$ where $\F$ appears in the last equation. This gives
\begin{widetext}
\begin{eqnarray} \nonumber
\frac{1}{\VE}
\W(E,t)
&=&\G(E,t)-(q-i)\frac{\Gamma\res}{2}\frac{i}{2\pi}\jnt{\omega'=-\infty}{+\infty}{
\frac{(2\pi)^{-1}
	\jnt{t'=-\infty}{+\infty}{
		\mathcal{E}(t')e^{i\omega' t'}
	}{t'}
}
{(\omega'-\omega\res+i\frac{\Gamma\res}{2})}
\lim\limits_{t_0\rightarrow-\infty}
\frac{\left[e^{i(\omega-\omega') t}-e^{i(\omega-\omega') t_0}\right]}{(\omega'-\omega)}
}{\omega'}\\ 
 \label{eqn:WEtK}
&=&\G(E,t)-(q-i)\frac{\Gamma\res}{2}\frac{i}{(2\pi)^2}\jnt{t'=-\infty}{+\infty}{
	\mathcal{E}(t')\left[e^{i\omega t}J(t'-t)-\lim\limits_{t_0\rightarrow-\infty}e^{i\omega t_0}J(t'-t_0)\right]
}{t'}
\end{eqnarray}
\end{widetext}
where we introduced 
\begin{eqnarray}
J(\tau)&:=&\intinf{\frac{e^{i\omega'\tau}}{(\omega'-\omega\res+i\frac{\Gamma\res}{2})(\omega'-\omega)}}{\omega'}.
\end{eqnarray}
Contour integrations give (see Appendix~\ref{sec:Kint})
\begin{eqnarray}
e^{i\omega t}J(t'-t)&=&-2\pi i e^{i\omega t}[1-\Theta(t'-t)] 
\nonumber \\ &&\times
\left[\frac{e^{i(\omega\res-i\frac{\Gamma\res}{2})(t'-t)}}{\omega-\omega\res+i\frac{\Gamma\res}{2}}-\frac{e^{i\omega(t'-t)}}{\omega-\omega\res+i\frac{\Gamma\res}{2}}\right]
\nonumber \\
\end{eqnarray}
where $\Theta$ is the Heaviside function, and
\begin{eqnarray}
\lim\limits_{t_0\rightarrow-\infty}e^{i\omega t_0}J(t'-t_0)&=&0.
\end{eqnarray}
\vspace{3cm}
Inserting this in Eq.~\ref{eqn:WEtK} leads to:
\begin{widetext}
\begin{eqnarray} \nonumber
\frac{1}{\VE}
\W(E,t)
&=&\G(E,t)-\frac{(q-i)}{(\varepsilon+i)}\frac{1}{2\pi}
\left
[
e^{i(\omega-\omega\res+i\frac{\Gamma\res}{2})t}
\underbrace{\jnt{t'=-\infty}{t}{
\mathcal{E}(t')e^{i(\omega\res-i\frac{\Gamma\res}{2})t'}
}{t'}}_{\G(E\res-i\frac{\Gamma\res}{2},t)}
-
\underbrace{\jnt{t'=-\infty}{t}{
\mathcal{E}(t')\,e^{i\omega t'}
}{t'}}_{\G(E,t)}
\right
]  \\ 
\end{eqnarray}
\end{widetext}
and we ultimately obtain the sought after expression for the \lift of $a(t)$,
\begin{eqnarray} \label{eqn:WEtFano} 
\W(E,t)&=&\VE\left[\frac{(\varepsilon+q)}{(\varepsilon+i)}\G(E,t)
\right. \nonumber \\ && \left. 
-\frac{(q-i)}{(\varepsilon+i)}e^{i\varepsilon\frac{\Gamma\res}{2}t}e^{-\frac{\Gamma\res}{2}t}\G(E\res-i\frac{\Gamma\res}{2},t)\right].
\end{eqnarray}
By comparing Eq.~\ref{eqn:WEtFano} with Eq.~\ref{eqn:cEtmerc}, one immediately sees that
\begin{eqnarray}\label{eqn:question}
\W(E,t)&=&ic_E(t).
\end{eqnarray}
This result shows that the \lift-based analysis of $A(E)$ gives access to a most fundamental quantity, the time-dependent spectral coefficients $c_E(t)$ of the continuum wave packet describing the photoelectron. From there, using the definition of $\PEt$ (Eq.~\ref{eqn:PEtdef}), we obtain Eq.~\ref{eqn:PEt}. The conjecture according to which the complete build-up dynamics of a Fano resonance can be retrieved from its spectral ionisation amplitude \A is therefore indeed valid. 

\subsection{
Numerical tests of the ``constant $V_E$'' approximation}\label{sec:numerics}

We have performed numerical simulations in order to illustrate the analytical results derived in the previous section, and to assess the validity of the ``constant $V_E$'' approximation. 

\subsubsection{Coupled-channel model}
We used a 1-dimensional coupled-channel model  designed to display an autoionising state with adjustable energy $E\res$, width $\Gamma\res$ and $q$ parameter. It is defined by a field-free hamiltonian $H_0$ explicitly partitioned as two \oned hamiltonians $h_1$ and $h_2$ associated with each of the channels coupled by a $V\cpl$ term,
\begin{eqnarray}\label{eqn:H_0}
H_0&=&\left(\begin{array}{ll} h_1 & V\cpl \\ V\cpl & h_2 \end{array}\right).
\end{eqnarray}
Each of the channel-specific hamiltonians $h_k$ ($k=1,2$) is assigned a standard form
\begin{eqnarray}
h_k&=&-\frac{1}{2}\frac{\partial^2}{\partial x^2}+V_k(x)
\end{eqnarray}
where $V_k(x)$ are potentials adapted to reproduce the desired energy spectra. In the present case, $V_1(x)$ is a gaussian potential whose parameters (depth and width) set the ionisation potential $I_p$ of the model atom. The continuum states of $h_1$ correspond to the \jeremie{reference scattering states}  \ket{\varphi_E}. The potential $V_2(x)$ was adjusted for $h_2$ to display a ground state at the same energy than $h_1$ and an excited {\em bound} state (with odd symmetry) at an energy located close to $E\res$, above the ionisation threshold of $h_1$. This excited state corresponds to \ket{\phi_{\mbox{\tiny b}}} in Eq.~\ref{eqn:psires}. The coupling term $V\cpl(x)$ is a hyper-gaussian function optimised empirically to obtain the desired resonance parameters, $q$ and $\Gamma\res$. This model reproduces very well the features of Fano resonances~\cite{yan2017a}, and is simple enough to allow extensive time-dependent and time-independent numerical simulations.

\begin{table}
\begin{tabular}{cccccc}
 atom & $I_p$ (eV) & $E\res$ (eV) & $\Gamma\res$ (meV) & $\tau\res$ (fs) & $q$ \\
\hline \hline
A & 24.59 & 35.62 & 39.2 & 16.67 & -2.98 \\
\hline
B & 15.53 & 1.54 & 57.7 &  11.32 & -2.66 \\
\hline
\end{tabular}
\caption{\label{tab:modelatoms} Characteristic features of the model atoms used in the simulations: ionisation potential ($I_p$), resonance energy  ($E\res$), width ($\Gamma\res$),  lifetime ($\tau\res=\Gamma\res^{-1}$) and Fano parameter ($q$). The energy scale refers to the ionisation threshold for each atom.}
\end{table}

Here we present results obtained with two such ``atoms'', with relevent properties summarised in table~\ref{tab:modelatoms}.
The first one (atom A) has characteristics close to the ones of He and its 2$s$2$p$ resonance~\cite{fano1961a}, which lies few tens of eV above threshold. The second one (atom B) displays an autoionising state with comparable width and $q$ parameter, but located very close to the threshold. For each atom, we simulated a resonant photoemission process by solving numerically the time-dependent Schr\"odinger equation (\tdse) in presence of a light pulse with adapted central frequency $\wX$ and full duration $\tauX$. The pulses were assigned $\sin^2$ envelopes centered at $t=0$, with intensities safely in the perturbative regime.

\subsubsection{Resonance far from threshold}
\begin{figure}
\center
\includegraphics[height=0.4\linewidth]{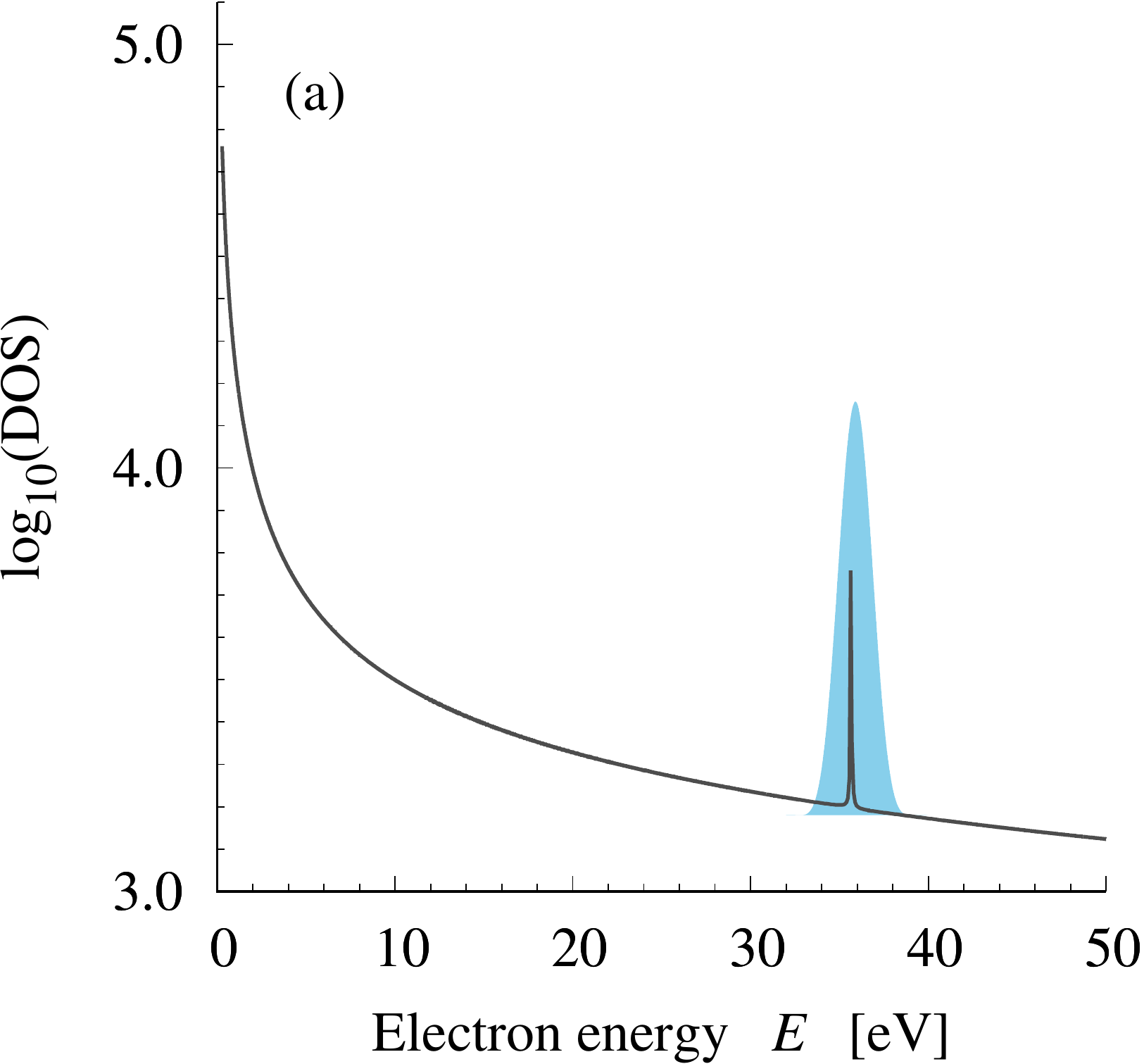}
\includegraphics[height=0.4\linewidth]{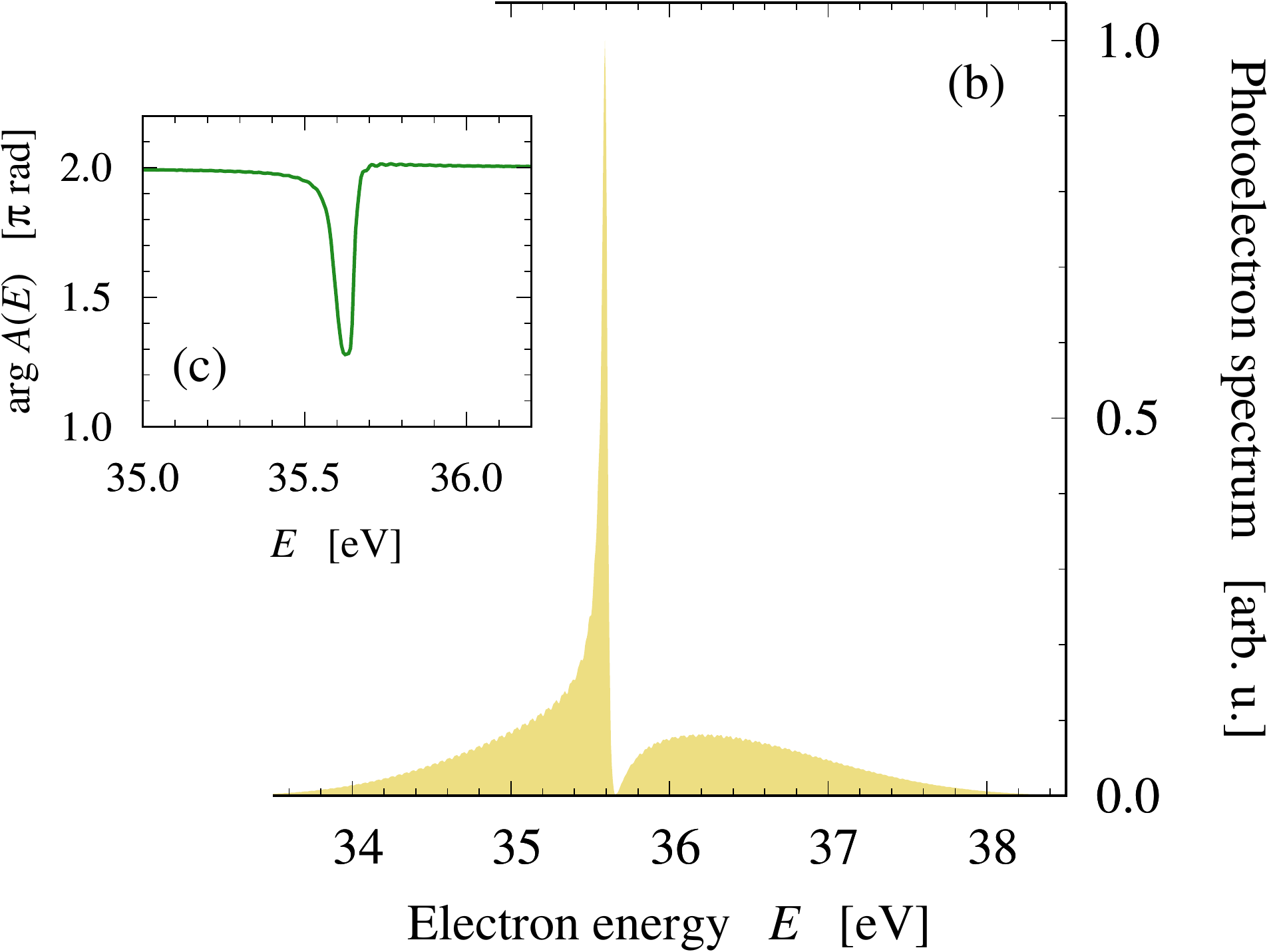}
\caption{\label{fig:DoS_A} (a) \dos\ above ionisation threshold for the  atom A (grey full curve)  and spectral envelope of the pulse used in the simulations (blue filled curve) ; (b) \PES $P(E)$ (yellow filled curve) ; (c) phase $\arg A(E)$ in the vicinity of the resonance (green full curve). The pulse parameters are $\wX=60.33$~eV and $\tauX=2.67$ fs.}
\end{figure}

We first detail our procedure and present the results obtained for atom A. The density of states (\dos) above ionization threshold for this atom is displayed in Fig.~\ref{fig:DoS_A}(a). It follows a typical decaying law $\sim(2E)^{-1/2}$ on top of which a sharp peak  indicates the position of the resonance slightly above 35 eV. We estimated the relative variations of $V_E$ in the vicinity of the resonance through
\begin{eqnarray}\label{eqn:Delta}
\Delta&=&\frac{\vert V_{E\res+\Gamma\res/2}-V_{E\res-\Gamma\res/2}\vert }{V_{E_R}\Gamma\res}.
\end{eqnarray}
For the atom A, it amounts to $4.3\times10^{-2}$ eV$^{-1}$. We also displayed in Fig.~\ref{fig:DoS_A}(a) the profile of the pulse used in the simulation (blue filled curve). Its central frequency \wX was set to $60.33$~eV to reach the vicinity of the resonance, and its duration \tauX to $2.67$~fs (39 optical cycles) which is significantly smaller than the resonance lifetime $\tau\res=16.67$~fs. Figure~\ref{fig:DoS_A}(b) shows the photoelectron spectrum (\PES) $P(E)$, computed from the propagated wave function $\psi(x,t\fin)$ at the end of the simulation ($t\fin\simeq130$ fs) using the window method~\cite{kulander1992a} detailed in Appendix~\ref{sec:numPES}. Its asymetric shape is consistent with the resonance parameters, notably the maximum near $E\res$ ($\varepsilon=0$) and the minimum located at $E=E\res-q\Gamma\res/2$ ($\varepsilon=-q$), as well as with the pulse parameters (spectral width $\sim 4$ eV).

\begin{figure}
\center
\includegraphics[width=\linewidth]{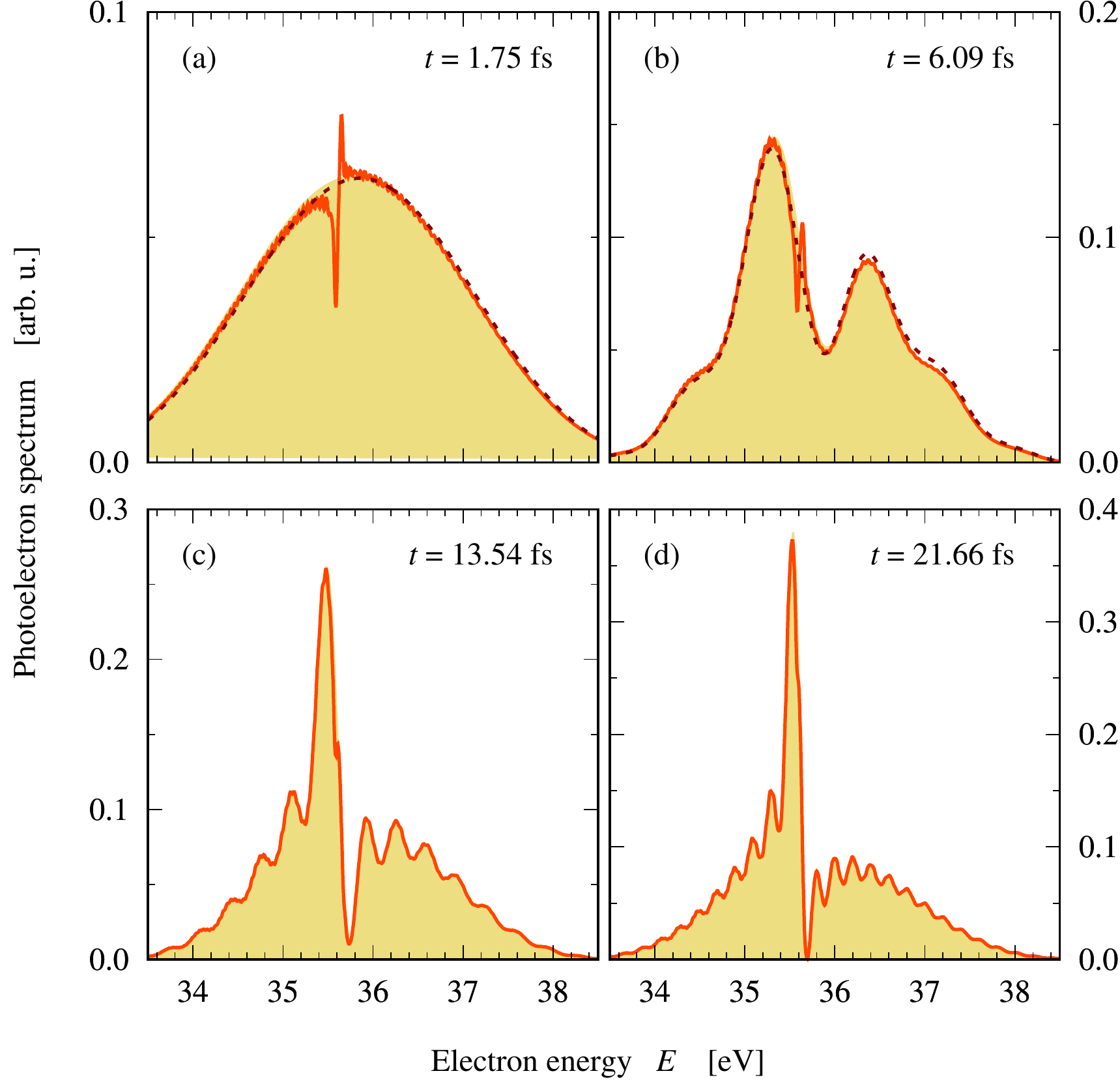}
\caption{\label{fig:PEtA} Transient photoelectron spectra for atom A  (pulse parameters specified in the caption of Fig.~\ref{fig:DoS_A}). (a)--(c): Actual \tPES, $\PEt$, (yellow filled curves) and reconstructed \tPES, $\vert \W(E,t)\vert^2$, (orange full curves) at four different times $t$. The time origin $t=0$ corresponds to the maximum of the pulse envelope. In frames (a) and (b) the  \tPES  reconstructed out of the analytical amplitudes \A is also shown (brown dashed curves).  }
\end{figure}

The  \tPES $\mathcal{P}(E,t)$ were computed using the same technique applied to the propagated wave-function $\psi(x,t)$ at various times during the process. The \tPES at four illustrative times are displayed in Fig.~\ref{fig:PEtA} (yellow filled curves). These numerically exact spectra are representative of the {\em actual} ionisation dynamics. Their evolution is consistent with the chronology of a Fano process~\cite{wickenhauser2005a} and notably with the analytical formula for the coefficients $c_E(t)$ (Eq.~\ref{eqn:cEtmerc}), as was already investigated in~\cite{mercouris2007a,chu2010a}. At the earliest displayed time, in frame (a), the spectrum follows a smooth bell curve reminiscent of the ionising pulse profile: at such an early time ($1.75\ \mbox{fs}\sim10\%\ \tau\res$), the ionisation process occurs dominantly through the {\em direct} path. In frame (b), first signatures of the autoionising path are already visible. The main peak begins to shrink around the resonance energy, on each side of which satellite structures appear. In frame (c), as the central peak continues to refine and to asymetrise, it becomes clearer that these satellite structures are oscillations, which result from interferences between the direct and autoionising paths. Indeed, it is easy to see from the analytical expression of the coefficients $c_E(t)$ in the case of a sudden pulse (see {\em eg} Eq.~23 of~\cite{chu2010a}),  that the \tPES contain an interference term proportional to $\sin [(E-E\res)t+\arctan q]$ -- the spectral frequency (phase) of which increases (decreases) linearly with time -- damped by a $\exp(-\Gamma\res t/2)$ factor. At the last displayed time in frame (d), the oscillations shrink and the spectrum converges to its final shape [see Fig.~\ref{fig:DoS_A} (b)].

Besides, still for atom A, we extracted the spectral amplitude $A(E)$ out of the final wave-function  $\psi(x,t\fin)$. The modulus is simply given as $\vert A(E)\vert=\sqrt{P(E)}$, while the phase $\arg A(E)$ was computed using an adapted numerical interferometric scheme detailed in Appendix~\ref{sec:numRABBIT}. The spectral variations of $A(E)$, displayed in Fig.~\ref{fig:DoS_A}(c), are typical of a Fano resonance with the two consecutive $\sim\pi$ jumps already discussed in Section~\ref{sec:Fanoparadigm}. From $A(E)$, we computed the temporal amplitude $a(t)$ (Eq.~\ref{eqn:E2t}) and its \lift $\W(E,t)$ (Eq.~\ref{eqn:WEtdef}). 
The reconstructed \tPES, $\vert \W(E,t) \vert^2$, are shown in Fig.~\ref{fig:PEtA}  (orange full curves), at the same times than $\PEt$ for comparison. Apart from a spurious narrow structure around $E\res$ visible at the earlier times [frames (a) and (b), hardly in frame (c)],  the agreement between the {\em actual} and the {\em reconstructed} spectra is excellent. We identified the spurious structure as a numerical artefact due to an imperfect convergence of the ``final'' wave-function $\psi(x,t\fin)$ used to compute $A(E)$. To verify this, we also computed the reconstructed $\tPES$ using the {\em analytical} expression for $\A$ (Eq.~\ref{eqn:AEFano}), with the Fano and pulse parameters adapted to our simulations, and displayed them in frames (a) and (b) of Fig.~\ref{fig:PEtA} (brown dashed curves). The spurious structure is indeed absent from these analytical reconstructed \tPES, which almost perfectly reproduce the actual \tPES (yellow filled curves). In conclusion, the results obtained with atom A illustrate and confirm the validity of Eq.~\ref{eqn:PEt} when $V_E$ is nearly constant.

\subsubsection{Resonance near threshold}
\begin{figure}
\center
\includegraphics[height=0.4\linewidth]{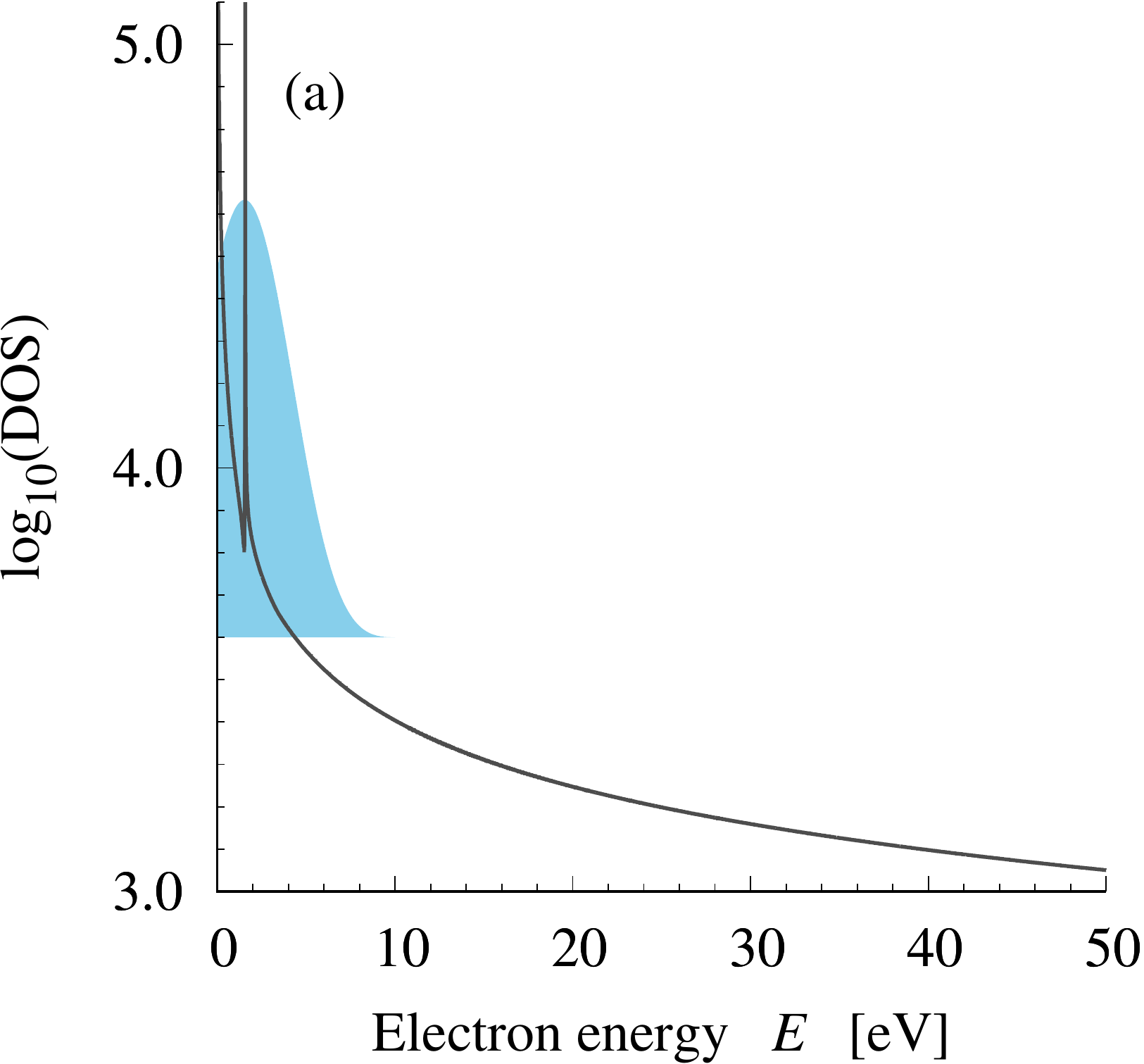}
\includegraphics[height=0.4\linewidth]{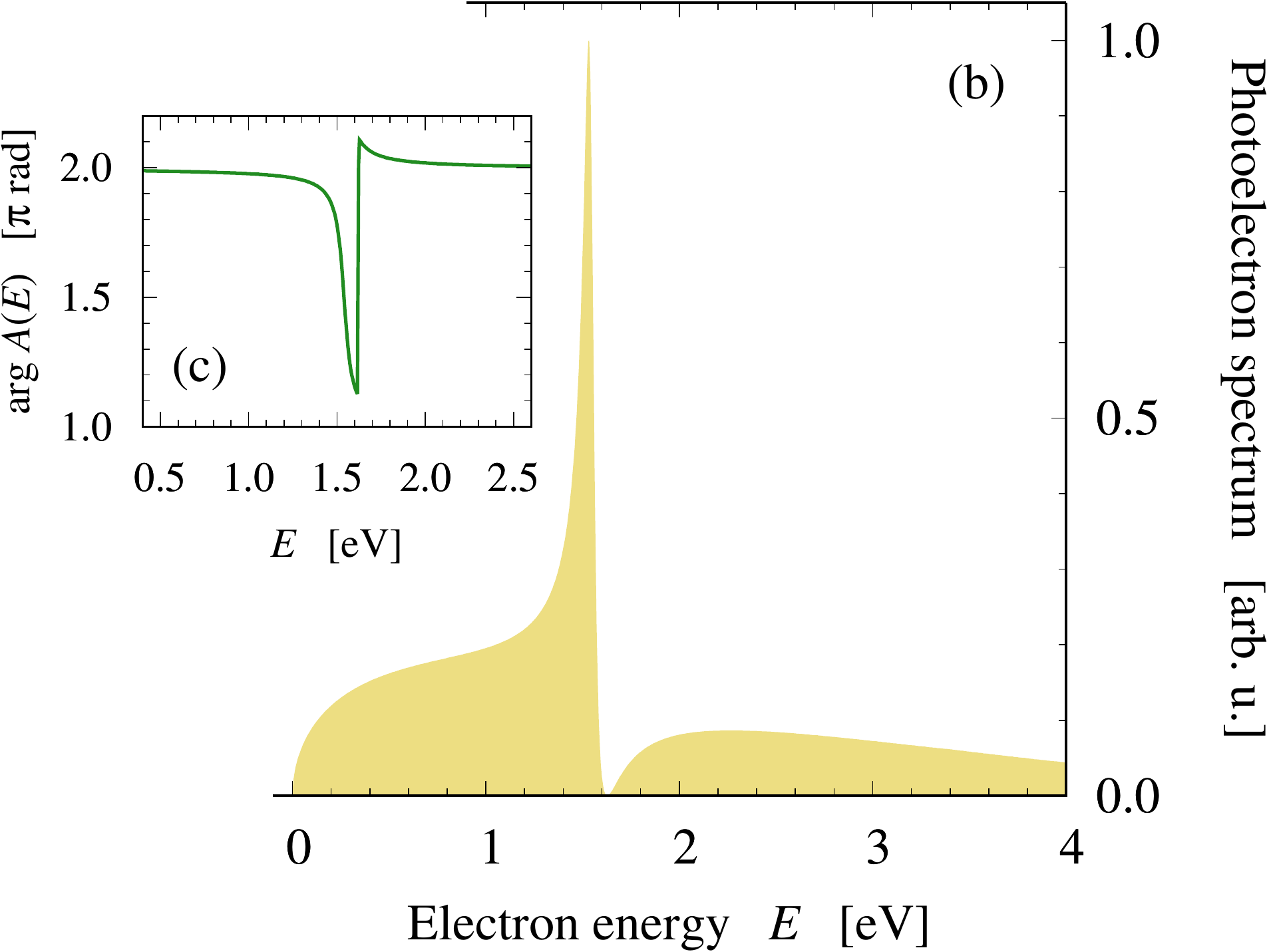}
\caption{\label{fig:DoS_B} (a) \dos\ above ionisation threshold for the  atom B (grey full curve)  and spectral envelope of the pulse used in the simulations (blue filled curve) ; (b) \PES (yellow filled curve) ; (c) phase $\arg A(E)$ in the vicinity of the resonance (green full curve). The pulse parameters are $\wX=17.06$ eV and $\tauX=0.97$ fs.}
\end{figure}

We repeated the procedure for atom B, which possesses a resonance much closer to the ionisation threshold and for which considering $V_E$ as constant is questionable. Here $\Delta=2.1\times10^{-1}$ eV$^{-1}$ (Eq.~\ref{eqn:Delta}), {\em ie} one order of magnitude larger than for atom A.  The \dos above threshold is shown in Fig.~\ref{fig:DoS_B}(a) together with the ionising pulse profile ($\wX=17.06$ eV and $\tauX=0.97$ fs, {\em i.e.} 4 cycles). We used a broader pulse than for atom A on purpose, in order to make it  significantly overlap the ionisation threshold. This strongly affects the final \PES (computed here at $t\fin\simeq250$~fs) which abruptly vanishes at $E=0$, see Fig.~\ref{fig:DoS_B}(b). The phase $\arg A(E)$ displayed in Fig.~\ref{fig:DoS_B}(c) is consistent with the resonance parameters. 

\begin{figure}
\center
\includegraphics[width=\linewidth]{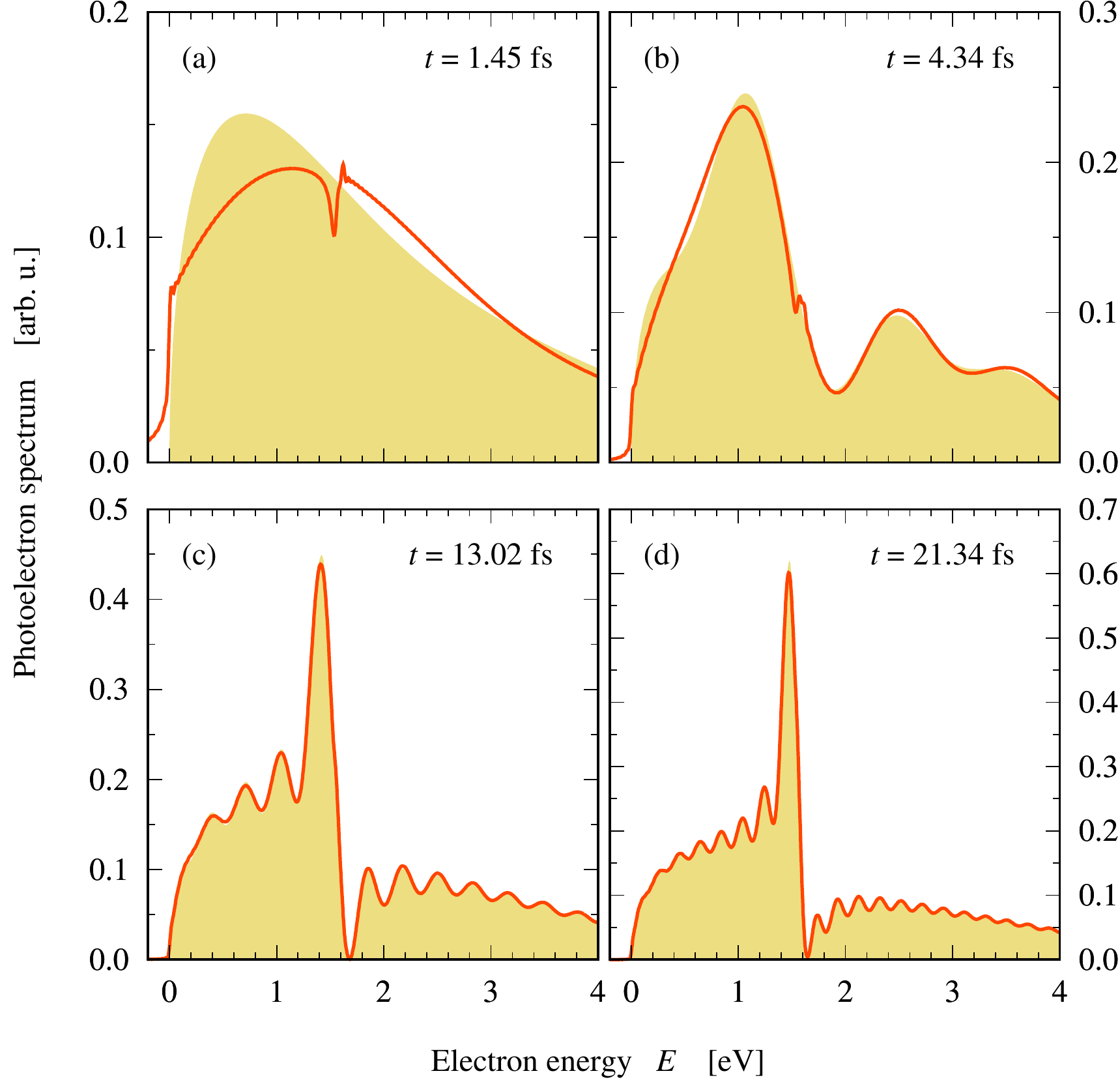}
\caption{\label{fig:PEtB} Transient photoelectron spectra for atom B  (pulse parameters specified in the caption of Fig.~\ref{fig:DoS_B}). (a)--(c): Actual \tPES, $\PEt$, (yellow filled curves) and reconstructed \tPES, $\vert \W(E,t)\vert^2$, (orange full curves) at four different times $t$. The time origin $t=0$ corresponds to the maximum of the pulse envelope.}
\end{figure}

The {\em actual} transient spectra $\PEt$ are displayed in Fig.~\ref{fig:PEtB} (yellow filled curves) at four times close to the ones chosen for atom A. Their evolution is similar to the one in atom A, notably the time-dependent spectral frequency of the transient oscillations, which does not depend on the resonance parameters. The two main differences with respect to atom A are: (i) the main peak building up over a somehow different time scale because of the different lifetime and (ii) a more pronounced asymmetry even at the earliest times [frame (a)] due to the non constant $V_E$ and to the above mentioned cut-off at $E=0$. 

The {\em reconstructed} ones, $\vert \W(E,t)\vert^2$, are overlaid on the same figure (orange full curves). As with atom A, they present a spurious narrow structure near $E\res$ at early times [frames (a) and (b)] resulting from an imperfectly converged $\psi(x,t\fin)$ used to compute them. Apart from this numerical artefact, it is clear  the agreement between $\vert \W(E,t)\vert^2$ and $\PEt$  at the earliest displayed time [frame (a)] is not as good as for atom A. This is a direct manifestation of non negligible variations of $V_E$ within the wave-packet support, which reshape its ``direct continuum'' components. Besides, the Fourier analysis does not perfectly reproduce the sharp cut-off at $E=0$. Thus the reconstructed \tpes in frames (a) and (b) slightly extend below the ionisation threshold. It is nevertheless noteworthy that a qualitative agreement between the shape and magnitude of $\PEt$ and $\vert \W(E,t)\vert^2$ remains in spite of particularly unfavourable conditions. Moreover, the excellent quantitative agreement between the reconstructed and actual \tPES is  recovered quickly, as can be seen already in frame (b) and confirmed in frames (c) and (d). Therefore, the results obtained with atom B seem to indicate that the ``constant $V_E$'' approximation,  needed to validate the conjectured Eq.~\ref{eqn:PEt}, is robust even for Fano resonances near threshold.

\subsection{Time domain interpretation of spectral amplitudes: General domain of validity}\label{sec:CGV}

The analytical and numerical results presented and discussed above  suggest to seek further for {\em general} conditions under which the temporal amplitude $a(t)$ and its \lift are physically relevent quantities.  To address this, we now step away from the Fano paradigm and simply consider a wave packet expressed as in Eq.~\ref{eqn:ewp}, where each \ket{\varphi_E} is the projection of an eigenstate of energy $E$ on a given reference partition/set. We also assume that the time-dependent coefficients of this wave packet, $c_E(t)$, and $A_E$ are related through  Eq.~\ref{eqn:AEcEt}.  The spectral amplitude $\A $ is in turn used to define the temporal amplitude $a(t)$ according to Eq.~\ref{eqn:E2t}. The wave packet has a finite norm, therefore $\A $   and its Fourier transform $a(t)$ are  square integrable functions. From there, we want to establish the condition(s) under which the \lift of $a(t)$, $\W(E,t)$, and the time-dependent coefficients, $c_E(t)$, verify Eq.~\ref{eqn:question}.

First, taking the time-derivative of Eq.~\ref{eqn:question}  with the definition of $\W(E,t)$ (Eq.~\ref{eqn:WEtdef}) gives
\begin{eqnarray}\label{eqn:82}
a(t)e^{iEt}&=&i\dot c_E(t)
\end{eqnarray}	
where we used the notation $\dot c_E(t)=\partial c_E(t) / \partial t$. A {\em necessary} condition for Eq.~\ref{eqn:82} (and therefore Eq.~\ref{eqn:question}) to be fulfilled, is that $\dot c_E(t)e^{-iEt}$ is independent of $E$, since $a(t)$ is by definition independent of $E$. 

We can moreover show that it is a {\em sufficient} condition, using the connections of $\A$ with $a(t)$ on the one hand, and with $c_E(t)$ on the other hand. Indeed, Eq.~\ref{eqn:E2t} implies
\begin{eqnarray}
\A&=&\frac{1}{2\pi}\intinf{a(t)e^{iEt}}{t} \label{eqn:81}
\end{eqnarray}
and Eq.~\ref{eqn:AEcEt} implies 
\begin{eqnarray}
\A&=&\frac{i}{2\pi}\lim\limits_{\tau\rightarrow+\infty}\jnt{-\infty}{\tau}{\dot c_E(t)}{t} \\
&=&  \frac{1}{2\pi}\intinf{\left[i\dot c_E(t)e^{-iEt}\right]e^{iEt}}{t}. \label{eqn:83}
\end{eqnarray}	
With that last expression, we see that if $i\dot c_E(t)e^{-iEt}$ does not depend on $E$, then it is the Fourier Transform of $\A$ -- just as $a(t)$ is according to Eq.~\ref{eqn:81}. Unicity of the Fourier Transform then implies Eq.~\ref{eqn:82}. 

Hence,  Eq.~\ref{eqn:question} is exactly fulfilled if and only if
\begin{eqnarray}\label{eqn:theorem}
\frac{\partial}{\partial E} \left[ \frac{\partial c_E(t)}{\partial t}e^{-iEt}\right]&=&0.
\end{eqnarray}
A physical interpretation of that condition is that the spectral components of the wave packet are homogeneously ``fed'' over time. Any structuring of the resulting wave packet is the consequence of interferences occurring during the build-up process. It is noteworthy that we  made here very little assumptions on the nature of the wave packet\footnote{It is nonetheless easy to verify that the coefficients $c_E(t)$ for a Fano resonance (Eq.~\ref{eqn:cEtmerc}) fulfill condition~\ref{eqn:theorem}.},  on the nature of the process populating the continuum states \ket{\varphi_E}, nor on the way the spectral amplitude is potentially measured experimentally. This is therefore a very general result. 

\subsection{Conclusions}\label{sec:conc}

In this article, we have established that the complete dynamics of a Fano autoionisation process could be retrieved out of the Fourier transform $a(t)$ of the associated transition amplitude \A\ -- a quantity that can be accessed experimentally under  conditions discussed elsewhere (see~\cite{gruson2016a} and references therein). The demonstration relies on a single approximation (a direct transition amplitude towards the continuum, $V_E$, which is spectrally constant), which is inherent to the Fano formalism and the validity of which is warranted by the broad applicability of the latter. It turns out that the limited inverse Fourier Transform of $a(t)$, $\W(E,t)$, exactly coincides, up to $i$ factor, to the time-dependent coefficient $c_E(t)$ of the quantum wave packet describing the photoelectron. The analytical demonstration was illustrated with simulations performed on simple model atoms displaying autoionising states, the results of which highlight the robustness of the ``constant $V_E$'' approximation.

Eventually, we have identified a general condition under which $\W(E,t)$  is equal to $ic_E(t)$ beyond the particular case of Fano resonances. This provides a unified framework within which the dynamics of a quantum wave packet can be retrieved out of its spectral scattering amplitudes \A through simple Fourier analysis. Our reasoning makes  little assumption on the nature of the wave packet and none on the way the spectral amplitudes are measured. It may therefore have broad implications ranging further than the particular case of a Fano resonance monitored in time by photoelectron interferometry~\cite{gruson2016a,beaulieu2017a,busto2018a} -- in a context where theoretical inputs are required not only to support experimental measurements, but most importantly for processing the measured data towards meaningful time-domain interpretations. 
\\
\begin{acknowledgements}
This project is supported by French state funds managed by the ANR programme ANR-15-CE30-0001-01-CIMBAAD and the LABEX Plas@Par-ANR-11-IDEX-0004-02.
\end{acknowledgements}

\appendix

\section{Analytics}

\subsection{How Eq.~\ref{eqn:PEt} implies Eq.~\ref{eqn:It}}\label{sec:demo1}

According to the conjectured Eq.~\ref{eqn:PEt}, the spectrally integrated ionisation probability at a given time $t$ reads
\begin{widetext}
\begin{eqnarray}
\intinf{\PEt}{E} &=& \frac{1}{(2\pi)^2}\intinf{\left\vert \jnt{-\infty}{t}{a(t')e^{iEt'}}{t'} \right\vert^2}{E}\\
&=&\frac{1}{(2\pi)^2}\jnt{-\infty}{t}{
\jnt{-\infty}{t}{
[a(t')]^\star a(t'')
\underbrace{\intinf{\left[e^{iE(t''-t')}\right]}{E}}_{\jeremie{2\pi}\delta(t''-t')}
}{t''}
}{t'} \\
&=&\frac{1}{\jeremie{2\pi}}
\jnt{-\infty}{t}{
\vert a(t') \vert^2 
}
{t'}.
\end{eqnarray}
\end{widetext}
Therefore,
\begin{eqnarray}
\frac{\partial}{\partial t}\intinf{\PEt}{E}&=&\frac{1}{\jeremie{2\pi}}\vert a(t) \vert^2.
\end{eqnarray}
Since the left-hand side of this last equation corresponds to the ionisation rate $I(t)$, we do obtain Eq.~\ref{eqn:It} as a consequence of Eq.~\ref{eqn:PEt}.

\subsection{Computation of the integral $J(\tau)$}\label{sec:Kint}

Equation~\ref{eqn:WEtK} involves the integral
\begin{eqnarray}
J(\tau)&=&\jnt{\omega'=-\infty}{\infty}{K(\omega',\tau)}{\omega'}
\end{eqnarray}
with
\begin{eqnarray}
K(\omega',\tau)&=&\frac{e^{i\omega'\tau}}{(\omega'-\omega\res+i\frac{\Gamma\res}{2})(\omega'-\omega)}.
\end{eqnarray}
It displays two poles at  $\omega$ and $\omega\res-i\frac{\Gamma\res}{2}$ respectively associated with the  residues
\begin{eqnarray}
\mathcal{R}(\tau)&=&\frac{e^{i\omega\tau}}{\omega-\omega\res+i\frac{\Gamma\res}{2}} ,\\
\mathcal{R}\res(\tau)&=&\frac{e^{i(\omega\res-i\frac{\Gamma\res}{2})\tau}}{\omega\res-i\frac{\Gamma\res}{2}-\omega}.
\end{eqnarray}
It can be evaluated in the complex plane as
\begin{eqnarray}\label{eqn:KCarc}
J(\tau)&=&\jnt{\mathcal{C}}{}{K(\omega',\tau)}{\omega'}-\jnt{\gamma}{}{K(\omega',\tau)}{\omega'}
\end{eqnarray}
where $\mathcal{C}$ is a contour consisting of the real axis {\em indented clockwise around the pole at $\omega$}, and closed by an arc $\gamma$ at infinity, either in the upper ($\uparrow$) or lower ($\downarrow$) half plane. The contours are such that the arc in the upper (resp. lower) half plane is followed counter-clockwise (resp. clockwise).   Choosing the most appropriate contour for the evaluation of $J(\tau)$ depends on the sign of $\tau$.

The full contour integrals are given by the residue theorem:
\begin{eqnarray}\label{eqn:Cupper}
\jnt{{\mathcal C}_\uparrow}{}{K(\omega',\tau)}{\omega'}&=&0
\end{eqnarray}
because there is no pole in the upper half-plane, and:
\begin{eqnarray}\label{eqn:Clower}
\jnt{{\mathcal C}_\downarrow}{}{K(\omega',\tau)}{\omega'}&=&-2i\pi\left[\mathcal{R}(\tau)+\mathcal{R}\res(\tau)\right]
\end{eqnarray}
since  the ${\mathcal C}_\downarrow$ contour encompasses the two poles, which are  circled clockwise. The integrals along the arcs depend on the sign of $\tau:$
\begin{itemize}
\item[-] When $\tau>0$,  the integral along the arc $\gamma_\uparrow$ vanishes. Relations~\ref{eqn:KCarc} and~\ref{eqn:Cupper} immediately give
\begin{eqnarray}
J(\tau)&=&0.
\end{eqnarray}
\item[-] When $\tau=0$, the integrals along both $\gamma_\uparrow$ and $\gamma_\downarrow$ vanish. Both contours can thus be used to show that
\begin{eqnarray}
J(0)&=&0.
\end{eqnarray}
\item[-] When $\tau<0$,  the integral along the arc $\gamma_\downarrow$ vanishes and we obtain
\begin{eqnarray}
J(\tau)&=&-2i\pi\left[\frac{e^{i(\omega\res-i\frac{\Gamma\res}{2})\tau}}{\omega-\omega\res-i\frac{\Gamma\res}{2}}-\frac{e^{i\omega\tau}}{\omega-\omega\res+i\frac{\Gamma\res}{2}}\right] .
\nonumber \\ 
\end{eqnarray}
\end{itemize}

\section{Numerics}

\subsection{Evaluation of the photoelectron spectra}\label{sec:numPES}
 
We computed numerically the \tPES with the window technique~\cite{kulander1992a} adapted to the present two-channel model. The spectrum at a given time $t$ is computed as
\begin{eqnarray}
\PEt&=&\intinf{\vert\chi_E(x,t)\vert^2}{x}
\end{eqnarray}
where 
\begin{eqnarray}
\chi_E(x,t)&=&W_E\psi(x,t)
\end{eqnarray}
is the propagated wave-function filtered with a narrow spectral window operator
\begin{eqnarray}\label{eqn:win}
W_E&=&\frac{\gamma^{2}}{(h_1-E)^{2}+i\gamma^{2}}.
\end{eqnarray}
The width of the window, $\gamma$, sets the numerical resolution of the spectra. Note that the expression of $W_E$ (Eq.~\ref{eqn:win}) involves the field free hamiltonian $h_1$, the positive energy spectrum of which consists in the continuum states \ket{\varphi_E} excluding the bound part of the resonance (see the partitioned hamiltonian $H_0$ given in Eq.~\ref{eqn:H_0}). At a sufficiently large time $t\fin$, the \tPES converges numerically to the \PES:
\begin{eqnarray}
P(E)&=&\mathcal{P}(E,t\fin).
\end{eqnarray}  
 
\subsection{Evaluation of the Fano phase}\label{sec:numRABBIT}
We retrieved the phase $\arg A(E)$ (displayed in Fig.~\ref{fig:DoS_A}(c) and Fig.~\ref{fig:DoS_B}(c) for atom A and B respectively)  out of the final wave-function $\psi(x,t\fin)$, using an interferometric scheme based on the same numerical tool than in Appendix~\ref{sec:numPES}. Consistently with the formalism of Fano and with the definition of $A(E)$, the  latter scheme takes for reference the wave-function $\psi\rfr(x,t\fin)$ propagated under the same conditions than $\psi(x,t\fin)$, but {\em without} the coupling term in the hamiltonian $H_0$ ($V\cpl=0$ in Eq.~\ref{eqn:H_0}). It  describes the direct ionisation path {\em only}. The spectral phase we are interested in is therefore the phase difference between the filtered wave-functions $\chi_E(x)=W_E\psi(x,t\fin)$ and $\chi_E\rfr(x)=W_E\psi\rfr(x,t\fin)$, where $W_E$ is the spectral filter defined in Eq.~\ref{eqn:win}. We thus coherently added them as
\begin{eqnarray}
\xi_E(x;\theta)&=&\chi_E(x)+\chi_E\rfr(x)e^{i\theta},
\end{eqnarray}
where $\theta$ is an arbitrary phase set manually, and fitted the signal $\vert \xi_E(x;\theta)\vert^2$  obtained for various values of $\theta$ with the generic function
\begin{eqnarray}
g(\theta)&=&A+B\times\cos(\theta+\eta).
\end{eqnarray}
For sufficiently large distances, the fitting parameter $\eta$ becomes independent of $x$ and converges to $\arg A(E)$. 
\bibliographystyle{ieeetr}
\bibliography{jcbib}

\end{document}